%% file: main.tex
\documentclass[sigconf, 9pt]{acmart}

\AtBeginDocument{%
  \providecommand\BibTeX{{%
    \normalfont B\kern-0.5em{\scshape i\kern-0.25em b}\kern-0.8em\TeX}}}
\copyrightyear{2025}
\acmYear{2025}
\setcopyright{rightsretained}
\acmConference[ISPD '25]{International Symposium on Physical Design}{March 16--19, 2025}{Austin, USA}
\acmBooktitle{International Symposium on Physical Design
(ISPD '25), March 16--19, 2025, Austin, USA}
\acmDOI{}
\acmISBN{}

\usepackage{graphics}
\usepackage{graphicx}
\usepackage{amsfonts}
\usepackage{amsmath}
\usepackage{multirow}
\usepackage{xcolor}
\usepackage{lipsum}
\usepackage{subfig}
\usepackage{soul}
\soulregister\cite7
\soulregister\ref7
\soulregister\pageref7
\usepackage{algorithm}
\usepackage{algpseudocode}
\algtext*{EndWhile}
\algtext*{EndIf}
\algtext*{EndFor}
\algtext*{EndProcedure}
\usepackage[super]{nth}
\usepackage[normalem]{ulem}

\makeatletter
\def\@mkbibcitation{\relax}
\makeatother

\copyrightyear{2025}
\acmYear{2025}
\acmConference[ISPD '25]{Proceedings of the 2025 International Symposium on Physical Design}{March 16--19, 2025}{Austin, TX, USA}
\acmBooktitle{Proceedings of the 2025 International Symposium on Physical Design (ISPD '25), March 16--19, 2025, Austin, TX, USA}
\acmDOI{10.1145/3698364.3705358}
\acmISBN{979-8-4007-1293-7/25/03}

\begin{document}

\title{
ML-QLS: Multilevel Quantum Layout Synthesis
}

\author{Wan-Hsuan Lin}
\email{wanhsuanlin@ucla.edu} \orcid{0000-0002-7486-2143}
\affiliation{
  \institution{University of California, Los Angeles}
  \city{Los Angeles}
  \state{CA 90095}
  \country{USA}}
\author{Jason Cong}
\email{cong@cs.ucla.edu} \orcid{0000-0003-2887-6963}
\affiliation{
  \institution{University of California, Los Angeles}
  \city{Los Angeles}
  \state{CA 90095}
  \country{USA}}
\renewcommand{\shortauthors}{Lin and Cong}

\begin{abstract}
Quantum Layout Synthesis (QLS) plays a crucial role in optimizing quantum circuit execution on physical quantum devices.
As we enter the era where quantum computers have hundreds of qubits, 
optimal OLS tools face scalability issues, 
while heuristic methods suffer significant optimality gap due to the lack of global optimization.
To address these challenges, we introduce a multilevel framework, which is an effective methodology for solving large-scale problems in VLSI design. 
In this paper, we present ML-QLS, the first multilevel quantum layout tool with a scalable refinement operation integrated with novel cost functions and clustering strategies. 
Our clustering provides valuable insights into generating a proper problem approximation for quantum circuits and devices. 
The experimental results demonstrate that ML-QLS can scale up to problems involving hundreds of qubits and achieve a remarkable 69\% performance improvement over leading heuristic QLS tools for large circuits, which underscores the effectiveness of multilevel frameworks in quantum applications.
\end{abstract}

\maketitle


\input{section/0_introduction}
\input{section/1_background}

\input{section/2_method}
\input{section/4_evaluation}
\input{section/6_conclusion}


\bibliographystyle{ACM-Reference-Format}
\bibliography{references}

\end{document}

%% file: section/0_introduction.tex
\section{Introduction}

Quantum computing has attracted immense research interest due to its exponential speedup for classical intractable problems.
Among various qubit technologies, superconducting qubits are one of the most promising platforms to realize large-scale quantum computing~\cite{ibm_quantum,rigetti, arute2019quantum}.
Within superconducting quantum processors, qubit connectivity is limited, meaning not all pairs of physical qubits are capable of performing a two-qubit gate. 
However, in the circuit, two-qubit gates can occur between any pair of program qubits.
To overcome the connectivity limitation, quantum layout synthesis (QLS) accommodates circuit connectivity to the hardware by introducing additional gates.

As the quantum processor is subject to noise, the solution quality of QLS is an important factor for the circuit performance.
First, with short coherence time, qubits will lose information without computation.
Thus, minimizing circuit depth is crucial to ensure successful information retrieval. 
Second, since gate operations are not perfect, additional gates introduced by QLS may exacerbate errors and prolong the circuit execution time.
Therefore, to utilize the full computation power of the quantum device, we should minimize the number of extra gates and circuit depth during QLS.

QLS has been proven to be NP-hard~\cite{siraichi_qubitallocation_2018,tan2020queko}.
Some existing efforts are heuristic ~\cite{siraichi_qubitallocation_2018,ho2018cirq,zulehner2018mapping_to_ibm_qx,web18-ibm-qiskit, zulehner_efficient_2019, siraichi_qubit_2019, li_sabre_2019,murali_formal_2019,sivarajah_tket_2020,kole_improved_2020,liu_notallswaparethesame_2022,wu_robust_2022,fan_QLSML_2022,huang2022reinforcement,park2022fsqm,huang2024ctqr,huang2024dear} to solve the problem efficiently. 
There also has been works on exact QLS tools, which often cast QLS problems into constraint programming problems
and rely on existing solvers to yield solutions~\cite{wille2014optimal,bhattacharjee2019muqut,wille2019mapping,tan2020olsq, tan2021olsqga, zhang_time-optimal_2021, molavi2022satmap,nannicini2022optimal, lin2023olsq2}.
Comparing the solution quality between the leading heuristic tool Sabre~\cite{li_sabre_2019} and exact tool OLSQ2~\cite{lin2023olsq2}, Sabre has a 6-7$\times$ optimality gap, while OLSQ2 takes more than one day to compile a size with 36 qubits and 54 gates~\cite{cong2023lightning}.
As the leading heuristic tools exhibit large optimality gaps and the exact tools struggle with scalability, the demand for scalable and effective QLS tools has been greater in an era of quantum processors with hundreds of qubits~\cite{gambetta2023utility}.


\begin{figure}[t]
    \centering
    \includegraphics[width=\linewidth]{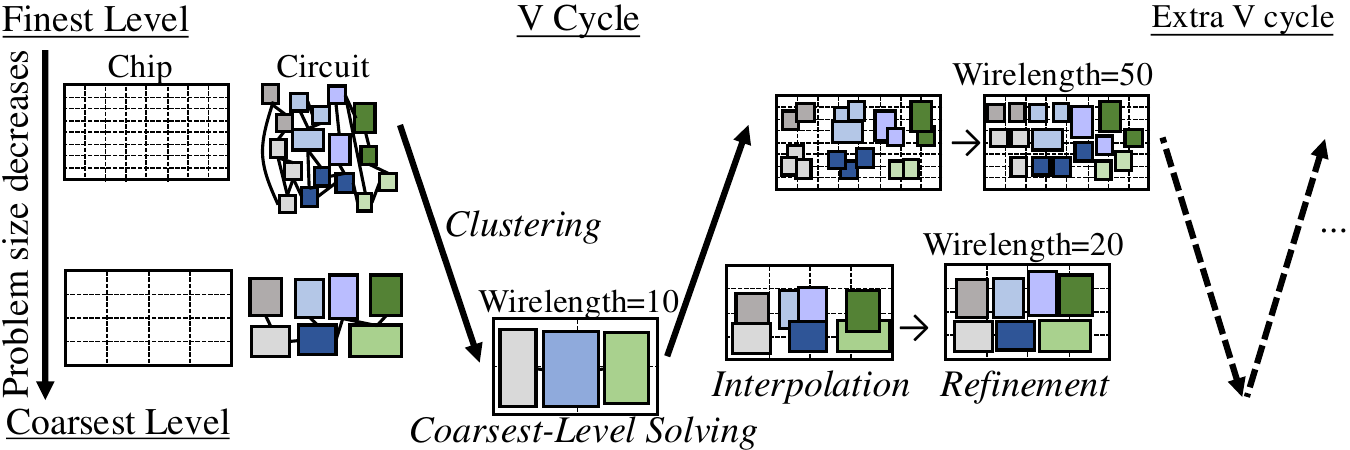}
    \caption{A multilevel V cycle for circuit placement.
    The inputs to circuit placement are a chip (a transparent box with black boundary) to place objects and a circuit consisting of placeable objects (colored boxes) and nets defining the connection between the objects.
    The process initiates with iterative clustering to reduce the problem size, continuing until reaching the coarsest level where the problem is directly solved. 
    The placeable objects marked in the same types of color form a coarser object at the coarser level.
    The dotted lines in chip represents the resolution at the current level.
    Subsequently, interpolation and refinement are used at each finer level, ultimately yielding the finest-level solution.}
    \label{fig:mflow}
    \vspace{-1.5em}
\end{figure}

To overcome the scalability issue while keeping a good solution quality, we introduces a novel multilevel framework to harness the strengths of both exact and heuristic approaches.
By employing an exact method at the coarsest level for its optimality and integrating scalable heuristics for refinement guided by the optimal coarser solution, we aim to create a hybrid solution that achieves superior performance across a wide range of quantum circuits.

Multilevel frameworks have demonstrated remarkable efficacy in large-scale optimization problems across various domains.
For example, they are applied in very large-scale integration (VLSI) design,
e.g., circuit partitioning~\cite{alpert1997multilevel,cong2004edge},
placement~\cite{chan_enhanced_2003,chen2005ntuplace,chan2005multilevel,cheng2018replace,cheng2018replace,leong2009replace,chan2000multilevel,chan2005mpl6}, and
routing~\cite{karypis1997multilevel,cong2005thermal,ou2012non,lin2002novel,liu2020cugr}, 
to deal with millions of transistors.
Figure~\ref{fig:mflow} shows an example of a multilevel framework for circuit placement.

In this context, the multilevel framework emerges as a powerful tool to tackle increasing problem sizes by constructing a hierarchical problem structure through \emph{clustering} to accelerate the exploration of the solution space at coarser levels.
The common approach is to group neighboring solutions at finer levels into a unified representation at coarser levels.
At the coarsest level, the problem can be solved optimally, often employing solver-based methods, to identify a promising region for further exploration. 
Although this stage might require a longer runtime, the investment is justified due to the valuable guidance it provides for the following refinement stages.

When transiting to a finer level, \emph{interpolation} is performed to project a coarser-level solution to the finer-level solution space through declustering.
Then, to enhance the solution at the current level, \emph{refinement} is applied to explore the neighboring solution space by leveraging information obtained from the coarser level.
This strategy enables efficient exploration of the search space even when the problem size is large.
In addition, a multilevel framework can adopt various multilevel flows, as illustrated in Figure~\ref{fig:mflow}.
The integration of multiple V cycles plays a crucial role in enhancing the solution's overall quality by allowing for the iterative refinement of clustering decisions.

In this paper, we present ML-QLS, which is the first work to apply a multilevel framework to provide high-quality results for large-scale quantum circuits. 
While multilevel frameworks have been widely studied in other fields, applying this approach to solve QLS presents unique challenges due to the specific characteristics of the problem. 
We analyze these challenges and provide effective solutions to address them. 
Our contributions are as follows:

\begin{itemize}
\item Clustering on discrete coupling graph posses a unique challenge to generate high-quality problem approximations. Thus, our clustering method leverages circuit clustering to guide device clustering to optimize the clustering decision.
\item We introduce sRefine, a heuristic tool that functions both as a standalone layout synthesizer for initial solutions and as a refinement operation within the multilevel flow.
With the concept of qubit region, sRefine effectively utilize the information from coarser level solution when serving as a refinement operation and demonstrates significant performance improvements by incorporating a qubit interaction cost term, previously overlooked by other methods.
\item ML-QLS reveals its effectiveness with 69\% SWAP reduction compared to the leading heuristic QLS tool. 
\end{itemize}

This paper introduces the multilevel framework to the quantum community, showcasing significant performance improvements in QLS and highlighting its potential applicability to a broader range of challenges in quantum computing design automation, e.g., qubit frequency calibration and compilation for other qubit platforms.
Through our comprehensive evaluation, we demonstrate that ML-QLS not only addresses the scalability issue but also maintains high solution quality, marking a significant advancement in this field.


%% file: section/1_background.tex
\section{Quantum Layout Synthesis}
\label{subsec:qls}
\begin{figure}[t]
    \centering
    \includegraphics[width=0.85\linewidth]{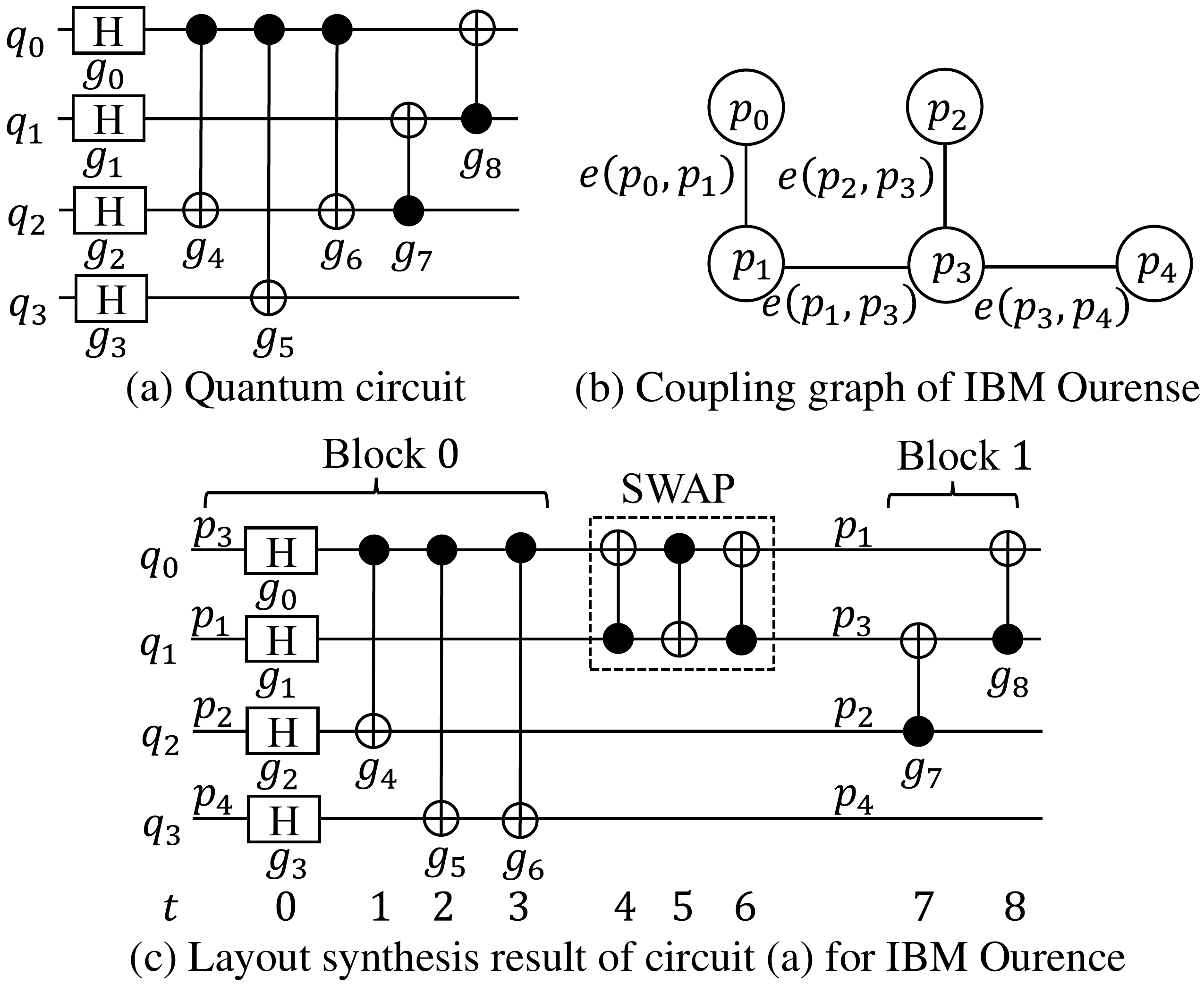}
    \caption{An example of a quantum circuit, coupling graph, and the corresponding QLS result.}
    \label{fig:example_circuit_graph}
    \vspace{-10pt}
\end{figure}



QLS is a process to map gates in a quantum circuit to a quantum processor defined by a coupling graph and transform circuit connectivity via inserting SWAP gates. 
The terminology for the inputs is defined as follows.
\paragraph{Quantum circuit:} A quantum circuit is defined by a sequence of gates with their target program qubits. 
In this paper, we denote the set of program qubits by $Q$, the set of single-qubit gates by $G_1$, the set of two-qubit gates by $G_2$, and the overall gate set by $G=G_1\cup G_2$. 
In addition, we refer to a gate on qubit $q$ and $q'$ by $g(q,q')$. 
Figure~\ref{fig:example_circuit_graph}(a) depicts a quantum circuit, where each horizontal line is a program qubit, H gates are single-qubit gates, and CNOT gates are two-qubit gates.

\noindent\paragraph{Coupling graph:} 
A coupling graph $(P,E)$ defines the connections between physical qubits. 
Each physical qubit is a vertex $p\in P$, 
and two physical qubits $p, p'$ can perform a two-qubit gate only if they are the endpoints of an edge.
In this paper, an edge between $p$ and $p'$ is denoted by $e(p,p')$. 
Additionally, we define a distance function $d: P\times P\rightarrow \mathbb{N}$, which is the distance of two physical qubits on the graph.
Figure~\ref{fig:example_circuit_graph}(b) illustrates a coupling graph of IBM Ourense~\cite{ibm_quantum}.

The output to the QLS problem consists of a mapping from program qubits to physical qubits, a gate schedule to indicate the gate execution time, and a list of inserted SWAP.
Figure~\ref{fig:example_circuit_graph}(c) shows a valid QLS result using one SWAP to map the circuit in Figure~\ref{fig:example_circuit_graph}(a) to IBM Ourense.
The physical location of a program qubit is indicated by the symbol next to the qubit line.
For example, program qubit $q_0$ is mapped to physical qubit $p_3$.

Due to the limited connectivity, one fixed mapping may not enable all gate execution.
Thus, SWAP gates are inserted into the circuit to adjust qubit mapping.
Having additional gates harms the circuit fidelity due to imperfect gate operations.
Thus, the objective of QLS is to minimize the number of inserted SWAP gates.
Here, we introduce the concept of blocks, where the qubit mapping remains the same.
For example, Figure~\ref{fig:example_circuit_graph}(c) consists of two blocks.

A valid QLS result should satisfy the following constraints:
\begin{enumerate}
    \item Mapping injectivity: At any time step, each program qubit should be mapped to a distinct physical qubit.
    \item Gate dependency: Non-commutable gates should be executed in order if they operate on the same qubits. 
    For example, in Figure~\ref{fig:example_circuit_graph}(a), since $g_0$ appears before $g_4$ in the gate sequence and they both act on qubit $q_0$,  $g_0$ should be executed before $g_4$.
    On the other hand, for circuits whose gates are commutable, the gates can be executed in any order.
    \item Valid two-qubit gate execution: To execute a two-qubit gate, its target program qubits should be mapped to adjacent physical qubits.
    For instance, in Figure~\ref{fig:example_circuit_graph}(c), $g_4$ can be executed because its target qubits $q_0$ and $q_2$ is mapped to an adjacent pair of physical qubits $p_3$ and $p_2$.
    \item SWAP transformation: The mapping transformation should be consistent with SWAP insertion.
    In Figure~\ref{fig:example_circuit_graph}(c), initially, $q_0$ is mapped to $p_3$ and $q_1$ is mapped to $p_1$. 
    After the SWAP gate occurs, $q_0$ is mapped to $p_1$, and $q_1$ is mapped to $p_3$. 
\end{enumerate}

%% file: section/2_method.tex
\section{Multilevel Quantum Layout Synthesis}
\label{sec:mOLSQ}

\begin{figure}[t]
    \centering
    \includegraphics[width=\linewidth]{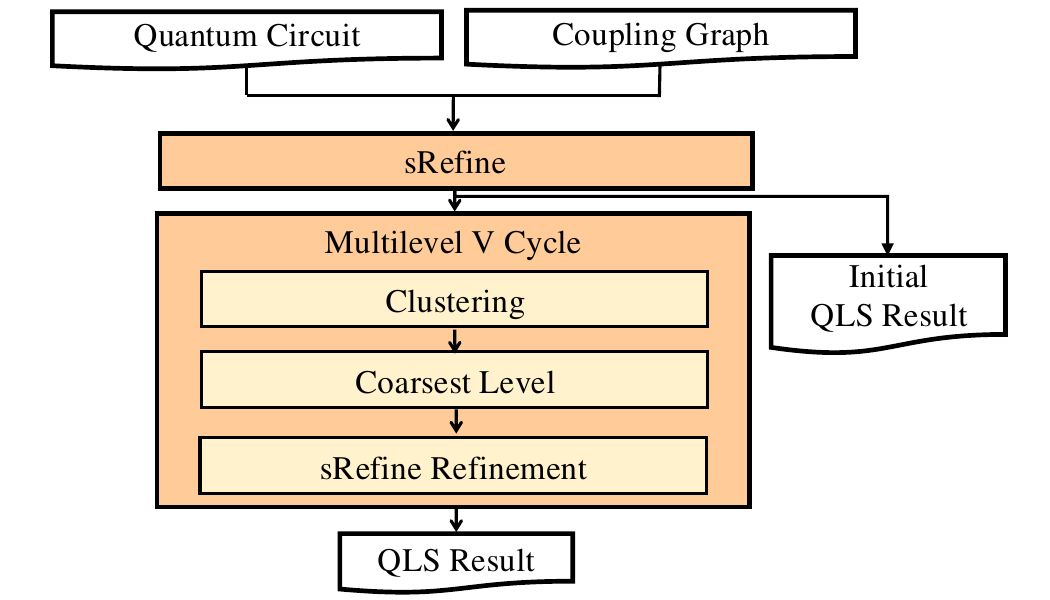}
    \caption{ML-QLS flow.}
    \label{fig:ML-QLS_flow}
    \vspace{-12pt}
\end{figure}

In this section, we offer a multilevel QLS tool, ML-QLS, consisting of two stages: initial QLS solution construction, followed by the multilevel V cycle.
Both stages provide a QLS solution.
Figure~\ref{fig:ML-QLS_flow} illustrates our overall workflow.
We first discuss the challenges of applying a multilevel framework to solve a QLS problem and our solutions in Section~\ref{subsec:overview}.
Then, in Section~\ref{subsec:clustering} to Section~\ref{subsec:coarser_level}, we detail our clustering and the coarsest-level solving techniques.
In Section~\ref{subsec:heuristic_alg}, we present sRefine, which is our scalable refinement algorithm.
Section~\ref{subsec:standalone_tool} details the flow to generate clustering reference in the first stage via sRefine.
Section~\ref{sec:scalability_analysis} discusses the scalability of ML-QLS.

\subsection{Multilevel Framework for QLS}
\label{subsec:overview}
Although the multilevel method is intuitive and well-developed in other domains, there are multiple challenges when applying to solve QLS problem.

\begin{figure}[t]
    \centering
    \includegraphics[width=0.95\linewidth]{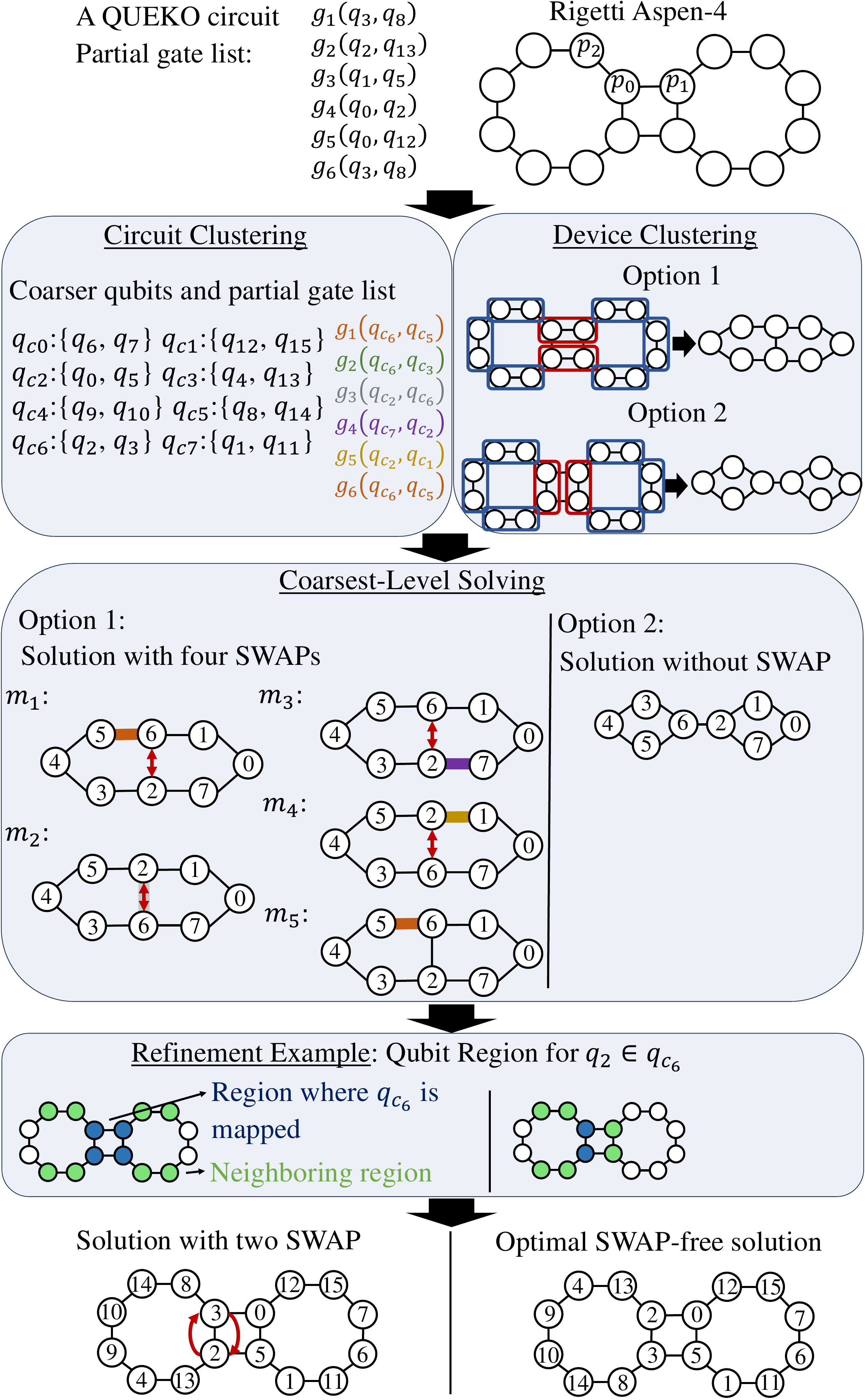}
    \caption{A V cycle example for a QUEKO circuit and Rigetti Aspen-4 coupling graph to demonstrate the effect of clustering on the final QLS solution.
    The partial gate list of the QUEKO circuit is shown at the top of the figure.
    According to the qubit interaction frequency, we generate a proper circuit clustering, including coarser qubits and the coarser partial gate list.
    A coarser qubit $q_{c}:\{q, q'\}$ indicates finer qubit $q$ and $q'$ form coarser qubit $q_{c}$.
    For device clustering, we may have two options to generate different coarser coupling graphs.
    At the coarsest level, the first option leads to four SWAP gates even if its later compilation result is optimal, while we get a SWAP-free solution with the second clustering option.
    $m_{i}$ denotes the $i$-th qubit mapping in the circuit,
    and the gate execution on an edge is marked by a thick line in $m_{i}$ using the same color for font in the coarser gate list.
    The edges for SWAP gates are indicated by red double arrow lines.
    With the coarsest-level solution, we show the qubit region for $q_{2}$ as a refinement example.
    After refinement, we obtain a solution using two and zero SWAP gates with the first and second clustering options, respectively.}
    \label{fig:running_exaemple}
    \vspace{-1.5em}
\end{figure}

\vspace{-1.5mm}
\subsubsection{Challenge 1 -- Difficulty in device clustering}
\label{subsubsec:challenge1}
In QLS, clustering is performed on both inputs to effectively reduce the problem size.
The device and circuit clustering refers to the clustering decision for the coupling graph and quantum circuit, respectively. 
One of the primary challenges in QLS arises from the nontrivial device clustering.
In traditional multilevel frameworks, clustering is typically applied independently to one problem input, such as the graph in graph partitioning or the circuit objects in VLSI placement. 
Even with the need to cluster the input device, i.e., chip, in VLSI placement or routing, the device clustering is trivial by applying different granularity to a chip.
For example, a chip with a dimension of 100$\times$100 can form a coarser chip with a dimension of 50$\times$50 by adjusting the unit grid length.
The trivial device clustering in VLSI placement does not alter the solution space because the coarser chips preserve the properties of the finest chip, e.g., the chip shape or the relative distance between points.
Therefore, the cost at the coarser level can serve as a good indicator of the finer-level result.

However, in QLS, the challenge arises from the clustering on a discrete coupling graph.
Thus, different clustering options define varied solution spaces as they generate distinct topology and alter the relative distance between points.
Thus, a proper coarser device should be selected by taking the input circuit into consideration.
Otherwise, suboptimal coarser-level solution may be produced to misguide the subsequent refinement.
For instance, Figure~\ref{fig:running_exaemple} illustrates two clustering options to construct coarser devices for Rigetti Aspen-4. 
The two coarser devices have distinct connectivity, and none of them has the same topology as Rigetti Aspen-4.
Additionally, the relative distance between two physical qubits changes as well.
In Rigetti Aspen-4, the distance between $p_0, p_2$ and $p_0, p_1$ is one, while the distance between $p_1, p_2$ is two.
With the coarser device generated by clustering option 1, the distance between $p_0, p_2$ remains one, but the distance between $p_0, p_1$ is changed to zero, and $p_1, p_2$ is changed to one. 
On the other hand, if we apply clustering option 2, the distance between those qubits remains the same. 
The relative distance change for qubits is unwanted as it alters the solution space by favoring certain QLS results.
As shown in Figure~\ref{fig:running_exaemple}, the optimal solutions at the coarsest level are inconsistent in terms of both SWAP count and qubit mapping with different device clustering, leading to different final QLS results after refinement.

In this example, option 2 is a better device clustering strategy as it produces solutions with lower SWAP count compared to option 1 at every stage.
According to the final QLS result from clustering option 2, we note that the circuit clustering is consistent with the device clustering. 
That is, the physical qubits are clustered iff the program qubits mapped to them are clustered.
Therefore, we propose a strategy that integrates information from both the device and the circuit to ensure harmonious alignment between them.
Our strategy is inspired by multilevel flow with multiple V cycles.
With such flow, the solution quality can be improved because the clustering decision will be continuously refined based on the former solution.
For instance, the cell clustering in the first V cycle for VLSI placement is often decided according to the circuit information.
When proceeding to the second V cycle, the clustering decision can take into account the cell physical location based on the placement solution derived from the first V cycle.
Thus, our clustering decisions are guided by a good solution obtained from our heuristic QLS tool, sRefine. 
For circuit clustering, we tend to cluster program qubits that have many two-qubit gate interactions and are mapped to adjacent physical qubits to take into account the physical information.
Then, the device clustering is derived based on circuit clustering and the QLS solution to ensure that device clustering decisions are informed by circuit information.

\vspace{-1.5mm}
\subsubsection{Challenge 2 -- Hidden cost within cluster}
\label{subsubsec:challenge2}
Clustering plays a pivotal role in generating problem approximations within the multilevel framework
as the cost within a cluster is often overlooked.
For example, in circuit placement, the wirelength of placeable objects within a cluster can not be observed at the coarse level.
Therefore, when we transit from a coarser level to a finer level, the cost often increases rather than decreases as shown in Figure~\ref{fig:mflow}.
Since the cost increase is hard to estimate, it is challenging to selecting an optimal solution at a coarser level.
QLS has the hidden cost problem as circuit placement as well because the cost of gate execution within a physical qubit cluster cannot be captured.
Thus, having an appropriate clustering strategy that effectively represents the original problem at coarser levels is important.

In QLS, the cluster size is a crucial factor for an accurate cost estimation.
While larger cluster sizes may reduce the number of levels in the cycle, they can introduce inaccuracies during the coarser-level solving.
In the context of program qubit clustering, we often cluster qubits that interact frequently, as these clusters remain physically close during circuit execution.
However, when choosing larger cluster sizes, a challenge emerges: Not all pairs of physical qubits within a cluster share direct connections, necessitating the insertion of hidden SWAP gates. 
This can lead to imprecise SWAP cost estimation, impacting solution selection during each stage. 
To address this, we encourage to have small clusters to ensure zero-cost gate execution within a cluster.

\vspace{-1.5mm}
\subsubsection{Challenge 3 -- Discrete solution space}
\label{subsubsec:challenge3}
Discrete solution space of QLS poses a significant challenge for interpolation and refinement.
For mathematical optimization problems where solutions transit smoothly from one level to another, people can easily project a coarser-level solution to the finer-level solution space by interpolation.
Then, the derived solution is the starting point for the refinement operation in the current stage.
In circuit placement, people can decide the location of objects based on the location of their corresponding coarser-level objects as depicted in Figure~\ref{fig:mflow} and gradually adjust the object location to reduce wirelength.
However, QLS involves discrete decisions such as qubit mapping, gate scheduling, and SWAP operations, making it impossible to directly derive an initial QLS solution based on a coarser-level solution.

For refinement, one of the challenges originates from the difficulty of recognizing a promising coarser-level solution for the finer level as discussed in Section~\ref{subsubsec:challenge2}.
In addition, minor changes in the coarse-level solution can lead to vastly different outcomes at the finer level, making it difficult to predict how optimizations will propagate across levels.
As the coarser-level solution may not be optimal for the finer-level problem, extracting useful information from the coarser-level solution to guide the finer-level solving is crucial.
In a multilevel framework, refinement operations are expected to perform local optimization and are encouraged not to excessively alter the initial solution from interpolation based on the assumption that the initial solution is globally optimized.
For example, in circuit placement, one common strategy is to add a term to minimize the objective displacement to the initial solution.

Unfortunately, such strategy does not work for QLS as we cannot derive an initial solution.
To overcome this issue,
we propose the concept of \textit{mapping regions}, which are promising solution regions suggested by the coarser-level solution. 
For each program qubit $q$, a mapping region $R_q$ is defined as follows: 
First, the physical qubits where $q$ is mapped are included in $R_q$. 
Then, to facilitate exploration of neighboring solution spaces, $R_q$ is expanded to encompass physical qubits within a one-hop distance from the mapped qubits. 
This expansion facilitates a more flexible and efficient exploration of adjacent solution options.
In case of misguidance from suboptimal solutions, our refinement operation encourages qubits to stay within their mapping regions but not restrict them, allowing for greater adaptability in the optimization process.
Figure~\ref{fig:running_exaemple} illustrates the construction of qubit regions, demonstrating how this approach permits effective exploration of neighboring solution spaces.

\vspace{-4mm}
\subsection{Clustering}
\label{subsec:clustering}
Given the QLS solution found by sRefine (to be described in Section~\ref{subsec:standalone_tool}), we initiate the clustering process by generating coarser program and physical qubits. 
For program qubits, we can define the affinity between qubit pairs as follows:
The affinity between two program qubits will increase by one for each two-qubit gate acting on them.
Next, according to the descending order of affinity values, we iteratively cluster two finer program qubits which are mapped to the adjacent physical qubits.
If any qubits remain unclustered, we group them with the adjacent cluster that has the smallest size.

For the physical qubits, we cluster them based on the corresponding program qubit clustering.
For instance, if program qubits $q$ and $q'$ are mapped to physical qubits $p$ and $p'$ respectively, while $q$ and $q'$ form a clustered program qubit, then $p$ and $p'$ will form a clustered physical qubit.
Then, we iteratively cluster an unclustered physical qubit with its unclustered neighbor or the smallest neighbor cluster.
For the coarser coupling graph, each cluster forms a coarser physical qubit.
We establish an edge between two coarser qubits $p_{c_i}$ and $p_{c_j}$ if there exists an edge between their finer qubits.

Regarding the construction of a coarser circuit, if a gate in the original circuit operates on two qubits belonging to distinct coarser qubits $q_c$ and $q_c'$, we include the gate $g(q_c, q_c')$ in the coarser circuit.
Conversely, if a gate operates on the same coarser qubit, we omit it and assumes that there is no inherent cost associated with executing a gate within a coarser physical qubit.
In addition, there is a dependency between two gates in the coarser circuit if there is a dependency between the corresponding gates in the finer circuit.
We repeatedly generate a coarser problem via clustering until the problem size is tractable for the coarsest-level solving.

\subsection{Coarsest-Level Optimization}
\label{subsec:coarser_level}
In the coarsest level, TB-OLSQ2~\cite{lin2023olsq2} is used to solve the problem.
TB-OLSQ2 is the state-of-the-art optimal SMT-based QLS tool and is scalable by adopting the concept of gate blocks to reduce the size of the SMT model compared to OLSQ2.
Their SMT formulation consists of three types of variables: 
(1) mapping variables to represent the mapping from program qubits to physical qubits for each block,
(2) gate time variables to encode the block for gate execution,
(3) SWAP variables to indicate the use of SWAP gate on each edge between gate blocks, 
and four types of constraints as discussed in Section~\ref{subsec:qls}.
Their SWAP optimization is a two-dimensional search along block number and SWAP count to generate Pareto optimal results.
In our implementation, we impose a runtime limit of 100 seconds after we obtain the first SWAP optimization results.
According to our experimental results, TB-OLSQ2 can return a solution for problem size with less 16 qubits and 50 gates within five minutes.
Thus, it can serve as a tool to provide high-quality solutions at the coarsest level for problems less than this threshold.

\subsection{Scalable Refinement: sRefine}
\label{subsec:heuristic_alg}
sRefine serves our scalable refinement operation that has two components: simulated annealing (SA)-based initial mapping and A*-based SWAP insertion.

\subsubsection{SA-Based Initial Mapping}
\label{subsubsec:sa_initial_mapping}
In this stage, our primary objective is to establish an initial mapping that facilitates the execution of all gates while minimizing the need for SWAP gates. 
To achieve this, we define a cost function as follows:
\begin{equation}
\label{eq:sa_cost}
\mathit{Cost}(m)= \mathit{DisForGates(m)} + \mathit{DisForRelatedQubits(m)},   
\end{equation}
where $m:Q\rightarrow P$ is the mapping from program qubits to physical qubits.
$\mathit{DisForGates}$ in Eq.~\ref{eq:sa_cost} accounts for the distance between target qubits for two-qubit gates, which is calculated by 
\begin{equation}
\label{eq:DisForGates}
\mathit{DisForGates}(m)=\sum_{g(q,q')\in G_2} w_g \times d(m(q),m(q')),    
\end{equation}
where $w_g$ is the weight of a gate.
To account for the fact that gates farther from the beginning of the circuit should have less influence on the initial mapping decision, $w_g$ decreases exponentially relative to the gate's distance from the beginning of the circuit.

\begin{figure}[t]
    \centering
    \includegraphics[width=0.95\linewidth]{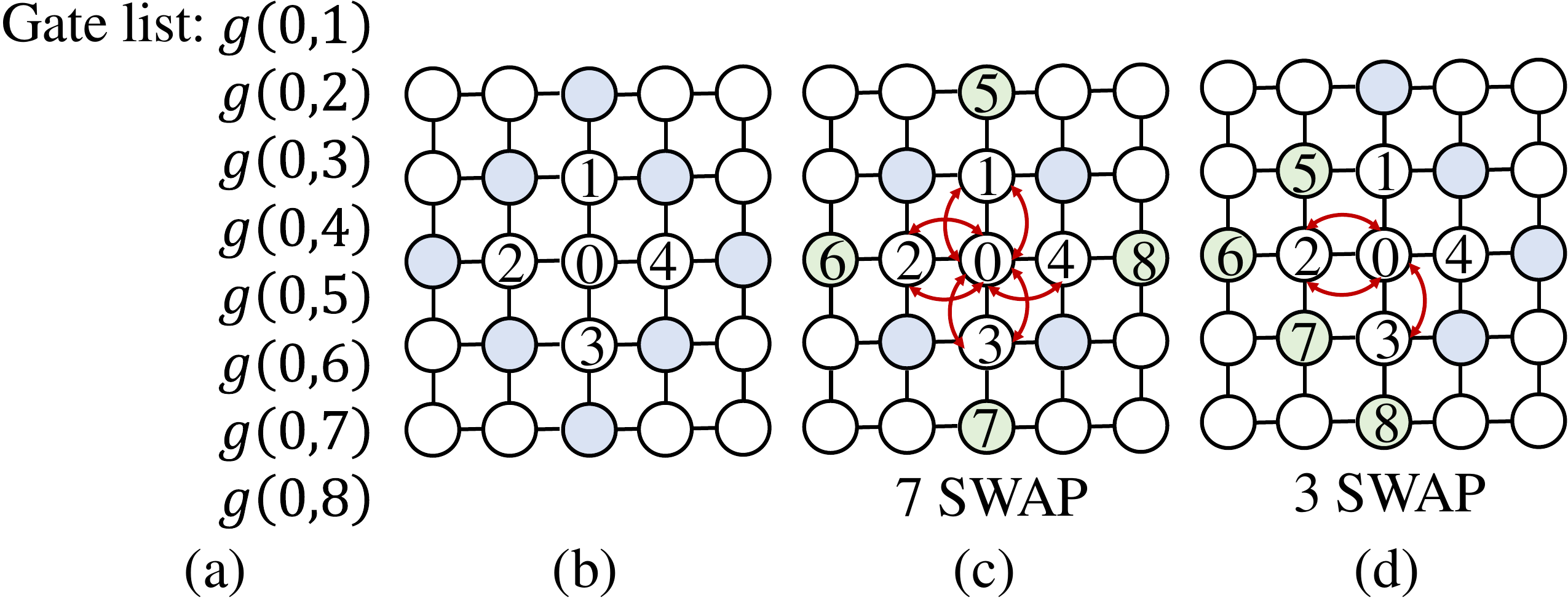}
    \caption{An example to exhibit the effectiveness of cost for related qubits.
    (a) A gate list. 
    (b) Optimal mapping on a grid coupling graph for $q_0$ to $q_4$ based on Eq.~\ref{eq:DisForGates}. 
    Placing $q_5$ to $q_8$ on any of the blue circles yields the same cost.
    (c) One optimal solution based on Eq.~\ref{eq:DisForGates} uses seven SWAP gates . 
    (d) One optimal solution based on Eq.~\ref{eq:sa_cost} employs three SWAP gates.}
    \label{fig:cost_example}
    \vspace{-15pt}
\end{figure}

Additionally, qubits that interact with the same qubit should sit near each other
such that one SWAP gate may enable multiple gate execution.
Figure~\ref{fig:cost_example} illustrates an example.
According to the gate list in Figure~\ref{fig:cost_example}(a), $q1$ to $q_7$ are related qubits as they all interact with $q_0$.
One optimal mapping for $q_0$ to $q_4$ under Eq.~\ref{eq:DisForGates} is shown in Figure~\ref{fig:cost_example}(b), 
and mapping $q_5$ to $q_8$ to any of the blue circles results in the same optimal cost. 
Figure~\ref{fig:cost_example}(c) demonstrates one solution that requires seven SWAP gates. 
In this example, each SWAP only enables at most one gate execution.
However, by arranging the related qubits nearby, as depicted in Figure~\ref{eq:DisForRelatedQubits}(d), we only need three SWAP gates, 
as the SWAP gate between $q_2$ and $q_0$ can enable three gate execution.
To capture such relation, we add the novel cost $\mathit{DisForRelatedQubits}$ to consider the distance between related qubits for two consecutive gates.
\begin{align}
\label{eq:DisForRelatedQubits}
\begin{split}
\mathit{DisFor}\mathit{RelatedQubits}(m)
=\sum_{\substack{g\in G_2 \\g'\in \mathit{Parent}(g)}} w_g \times d(m(q'),m(q'')),    
\end{split}
\end{align}
where $\mathit{Parent}(g)$ denotes the set of the last two-qubit gates acting on the target qubits of $g$ while $q'$ and $q''$ represent the uncommon qubits between $g$ and $g'$.
For instance, in Figure~\ref{fig:cost_example}, $g(0,2)\in \mathit{Parent}(g(0,3))$ because both gates operate on $q_0$, and $g(0,2)$ is the last previous two-qubit gate involving $q_0$.

During refinement, qubits have a higher chance to move within its qubit region derived from the coarser level solution, and the initial solution for SA is a maximal matching within the bipartite graph $G(V,E)$ derived from the coarser-level solution, where $V=Q\cup P$ and $(q,p)\in E$ if $p\in R_q$.
Note that the maximal matching can be obtained by Edmonds' Blossom algorithm in $O(|E||P|^2)$~\cite{edmonds1965paths}.

Subsequently, we utilize the 
SA algorithm~\cite{chen20006fastsa} to optimize our cost function. 
Within the SA-based initial mapping method, each state defines a unique mapping configuration. 
The neighbors of a state are generated by selecting a qubit and relocating it to a new position. 
If the chosen location is already occupied by another qubit, we exchange their positions.

\subsubsection{A*-Based SWAP Insertion}
\label{subsubsec:a_swap_insertion}

In contrast to the approach of inserting one SWAP gate at a time, as employed in Sabre~\cite{li_sabre_2019}, we adopt a strategy that selects multiple SWAP gates for a set of gates simultaneously to have more global information when deciding qubit movement.
Consequently, we develop an A*-based algorithm capable of evaluate the cost of multiple SWAP gates before committing a solution. 

In our A* search, a state 
$v$ consists of five elements:
\begin{itemize}
    \item An edge $e$ where the SWAP occurs, 
    \item A set of gates $G_{\mathit{ready}}$ ready for execution, i.e., their ancestors are all mapped, but cannot be executed under the current mapping, 
    \item A set of gates $G_{\mathit{unexecuted}}$ not yet ready for execution, 
    \item Its parent state $u$
    \item The cost $f(v)=g(v)+h(v)$, where $g(v)$ is the SWAP cost from the initial state to the current state, and $h(v)$ is an estimated SWAP cost from the current state to the goal state.
\end{itemize}
When transitioning from a state $u$ to its child state $v$, we alter the mapping by inserting one SWAP gate.
Consequently, we define $g(v_0) = 0$, where $v_0$ represents the initial state, and $g(v) = g(u) + 1$, where $u$ is the parent state of $v$.

The heuristic cost $h(v)$ is calculated as follows:
\begin{align}
\label{eq:h_cost}
\begin{split}
    h(v) = &\frac{\sum_{g(q,q')\in G_{\mathit{ready}}}{d(m(q),m(q'))}}{|G_{\mathit{ready}}|*|Q|} \\
    &+\alpha\times\frac{\sum_{g(q,q')\in G_{\mathit{oneHopReady}}}{d(m(q),m(q'))}}{|G_{\mathit{oneHopReady}}|*|Q|}\\ 
&+\beta\times\frac{\sum_{g'(q,q')\in\mathit{Parent}(g), g\in G_{\mathit{oneHopReady}}}{d(m(q),m(q'))}}{|G_{\mathit{oneHopReady}}|*|Q|}\\
    &+ \gamma\times|G_{\mathit{unexecuted}}|, \\ 
\end{split}
\end{align}
where $m(q)$ denotes the current mapping of $q'$, and $G_{\mathit{oneHopReady}}=\{g'|g'(q,q')\in\mathit{Child}(g), g\in G_{\mathit{ready}}\}$.
$\alpha,\ \beta,\ \gamma$ are the user-defined hyperparameters. 
In our implementation, we set $\alpha=\beta=0.5$ and $\gamma=0.1$.
The first term in the cost depicts the normalized qubit distance cost for the gates in $G_\mathit{ready}$, 
while the second term serves as the lookahead cost and represents the normalized qubit distance cost for the gates in $G_\mathit{oneHopReady}$.
The third term is the normalized cost for the related qubit distance as defined in Eq.~\ref{eq:DisForRelatedQubits}, 
and the last term signifies the number of unexecuted gates, capturing the cost for those gates not contributing to the distance cost.
The weight assigned to the last term determines the trade-off between SWAP insertion and gate execution. 
A weight of 0.1 signifies that it is worthwhile to insert one SWAP gate for every ten gate executions.

To address the state explosion problem inherent in A* search, we design a state trimming mechanism.
When the number of states exceeds a predefined threshold $s$, we retain only the $k$ states with the lowest cost. 
Empirically, we set $s=100$ and $k=50$.
In addition,
we restrict our consideration to SWAP gates associated with the target qubits of gates in $G_{\mathit{ready}}$.
Then, in alignment with Eq.~\ref{eq:h_cost}, we encourage target qubits to be moved closer to each other. 
Consequently, if a SWAP gate would relocate a qubit farther from its target qubit, we opt not to expand the node, thus conserving computational resources.
To follow the guidance from the coarser-level solution, the node associated with a SWAP that moves a qubit out of its qubit region will have a small probability of being expanded.
The search process terminates when all gates are executed. 
The QLS solution is obtained by backtracking the states to obtain the SWAP gates.
The gate scheduling can be derived via as-soon-as-possible scheduling.

In this stage, we refine the initial mapping based on routing information by applying the concept of forward and backward compilation passes, as proposed in~\cite{li_sabre_2019}. 
In our framework, we utilize the mapping returned by the SA-based initial mapping as the initial mapping for our first forward pass. 
This iterative process continues until the SWAP count from the current compilation pass is not better than the previous one. 
The best result obtained thus far is considered our final solution at the current stage.
This approach leverages the interplay between forward and backward compilation passes to iteratively refine the mapping, ultimately yielding an optimized solution.

\subsection{Initial Flow for Clustering Generation}
\label{subsec:standalone_tool}
sRefine can serve as a standalone tool to generate clustering guidance in the first stage by removing the qubit region restriction for neighboring state generation in SA and node expansion in A* search.
Additionally, when used as a standalone QLS tool, we design InitialMapper to provide a good quality initial solution for the SA algorithm.
InitialMapper is an SMT-based method designed to generate a mapping that enables the execution of most gates within the circuit.
Although we can use a random mapping as the initial solution, a high quality starting point can faciliate efficient solution exploration for SA.

The formulation of InitialMapper is a simplified variant of TB-OLSQ2 by removing time dimension, SWAP, gate time variables and the related constraints.
InitialMapper only contains mapping variables $m(q)$ for each qubit and associated mapping constraints.
Our approach involves a stepwise addition of gates $g(q,q')\in G_2$ with the constraint
\begin{align}
\begin{split}
\bigwedge_{e(p,p')\in E} &((m(q)==p)\wedge(m(q')==p'))
\\&\vee((m(q')==p)\wedge(m(q)==p')).    
\end{split}
\end{align}

This constraint is incorporated into the solver, and the iterative process unfolds as follows: 
If the instance is satisfiable, we proceed to add the subsequent gate following the predefined gate order.
Conversely, if the instance turns out unsatisfiable, we remove the constraint associated with the most recently added gate and introduce the constraint for the subsequent gate in line. 
Throughout this process, we maintain the solution with the lowest distance cost as defined in Eq.~\ref{eq:sa_cost}.
The candidate with the lowest cost is then designated as the initial solution for the SA-based initial mapping.

Regarding the gate order, while one could use the na\"ive order derived from the circuit, such an approach may not maximize the number of executed gates and could favor solutions with fewer SWAP gates at the beginning of the circuit. 
Therefore, we opt to generate the gate order by shuffling gates randomly to mitigate potential biases toward certain types of solutions.

Since both InitialMapper and the SA-based initial mapping incorporate randomness, we generate five initial mapping candidates for the subsequent SWAP insertion stages to produce layout synthesis solutions. 
The best among these candidates is selected as the final result. 
To manage the exponential growth in SMT solving time, we only invoke InitialMapper if the number of program qubits is fewer than 100. 
Additionally, we impose a runtime limit of 1000 seconds for the first run and 100 seconds for subsequent runs.

\subsection{Scalability Analysis}
\label{sec:scalability_analysis}
The complexity of ML-QLS is primarily determined by sRefine and the stages in the multilevel framework. 
With a single V cycle and a compression rate of two, the multilevel framework consists of $O(\log n)$ levels, where $n$ is the number of qubits.
Although the coarsest level solving invokes an SMT solver, the complexity is constant since the problem size at the stage is fix, and we impose a runtime limit for the SMT solving.

At each level, sRefine is applied to solve the problem, and the finest-level solving, which handles the largest problem size, dominates the overall complexity. 
The SA-based initial mapping has a complexity of $O(n)$ for the initial cost computation, and with a fixed iteration limit, the exploration time remains constant. 
Additionally, the complexity of Edmond's Blossom algorithm~\cite{edmonds1965paths} for  initial solution generation in SA is $O(n^3)$.
For the $\text{A}^*$-based SWAP insertion, the complexity is $O(s^d)$, where $s$ is the state number threshold, and $d$ is the search depth, related to the number of SWAPs needed. 
Therefore, the overall complexity of ML-QLS is $O(n^3+s^d)$. 
While the $\text{A}^*$ search is exponential, its cost can be reduced by using a smaller threshold, allowing users to balance between solving time and solution quality.

%% file: section/4_evaluation.tex
\vspace{-3pt}
\section{Evaluation}
\label{sec:evaluation}

\vspace{-3pt}
\subsection{Experimental Settings}
\paragraph{Benchmark and Baseline}
Our benchmarks include 
1) QUEKO circuits~\cite{tan2020queko}, which are the circuits with known-optimal compilation results used for layout synthesizers evaluation, 
2) QAOA~\cite{zhou_quantum_2020} phase-splitting operator with commutable gates for random 3-regular graphs, 
and 3) circuits from QASMBench~\cite{li2022qasmbench}.
The QAOA graphs are generated by networkx (v2.4)~\cite{networkx}. 

We test on two types of coupling graphs: grid architectures and heavy-hexagon architectures.
Grid architectures includes the Google’s Sycamore processor with 54 qubits~\cite{arute2019quantum}, and 2D-nearest-neighbor architectures with grid lengths ranging from 6 to 25.
In addition, we use IBM’s Eagle processor with 127 qubits~\cite{chow2021ibmeagle} to represent the heavy-hexagon architecture. 
We compare the performance of our tool against the leading heuristic layout synthesizers, Sabre~\cite{li_sabre_2019} with Qiskit (v1.2.0) implementation, t$|$ket$\rangle$~\cite{tket}, MQT QMAP~\cite{wille2023mqt}.
For Sabre, we set the trial number to four.
Note that we do not include a comparison with tools that solely focus on SWAP insertion, e.g., DEAR~\cite{huang2024dear}, as ML-QLS addresses the full quantum layout synthesis process, offering a more comprehensive solution.
Additionally, we do not report the comparison with the optimal QLS tool OLSQ2~\cite{lin2023olsq2} as OLSQ2 cannot produce a solution for our benchmark circuits with a runtime limit of 8 hours.

\paragraph{Experimental Platform}
We implemented our proposed algorithm in C++ programming language with Boost C++ libraries~\cite{boost} and provided a Python interface via PyBind11~\cite{pybind11} for evaluation.
We employed the Bitwuzla (v0.4.0)~\cite{bitwuzla} for SMT solving and pblib~\cite{pblib} for cardinality constraint-to-CNF generation. 
All experiments were conducted on an AMD EPYC 7V13 64-Core Processor at 2450 MHz and 128 GB of RAM.

\begin{table}[]
\begin{tabular}{cllllll}
\hline 
\hline
\multicolumn{1}{l}{\multirow{2}{*}{Grid}} & \multicolumn{1}{c}{\multirow{2}{*}{Circuit}} & \multicolumn{1}{c}{\multirow{2}{*}{Sabre}} & \multicolumn{1}{c}{\multirow{2}{*}{t$|$ket$\rangle$}} & \multicolumn{1}{c}{\multirow{2}{*}{QMAP}} & \multicolumn{2}{c}{ML-QLS} \\
\multicolumn{1}{c}{} & \multicolumn{1}{c}{{*}} & \multicolumn{1}{c}{} & \multicolumn{1}{c}{} & \multicolumn{1}{c}{} & sRefine & V cycle \\ \hline
6 & adder\_28 & 51 & 70 & 87 & \textbf{32} & 33 \\
10 & adder\_64 & 169 & 211 & 222 & \textbf{80} & \textbf{80} \\
11 & adder\_118 & 414 & 425 & 474 & 317 & \textbf{218} \\
9 & cat\_65 & 54 & 2 & \textbf{0} & \textbf{0} & \textbf{0} \\
6 & dnn\_33 & 78 & 44 & 62 & 35 & \textbf{29} \\
7 & ghz\_40 & 21 & 2 & \textbf{0} & \textbf{0} & \textbf{0} \\
6 & isling\_34 & 10 & 19 & 25 & \textbf{0} & \textbf{0} \\
10 & isling\_98 & 75 & 69 & 81 & \textbf{0} & \textbf{0} \\
6 & knn\_31 & 33 & 44 & 42 & \textbf{26} & \textbf{26}  \\
9 & knn\_67 & 110 & 114 & 125 & 132 & \textbf{90} \\
6 & wstate\_33 & 17 & 4 & 6 & \textbf{0} & \textbf{0} \\ \hline
\multicolumn{2}{c}{Geo. Ratio} & 6.03 & 3.43 & 3.50 & 1.09 & \textbf{1.00} \\
\hline 
\hline
\end{tabular}
\vspace{2pt}
\caption{SWAP count comparison for circuits from QASMBench benchmark suites on grid-based coupling graph.
    The number after underscore denotes the number of program qubits in the circuit.    
    For example, adder\_28 represents the adder circuit with 28 qubits.}
\vspace{-18pt}
\label{tab:qasm-grid}
\end{table}


\vspace{-14pt}
\begin{table}[]
\begin{tabular}{cllllll}
\hline \hline
\multirow{2}{*}{Arch.} & \multicolumn{1}{c}{\multirow{2}{*}{Circuit}} & \multicolumn{1}{c}{\multirow{2}{*}{Sabre}} & \multicolumn{1}{c}{\multirow{2}{*}{t$|$ket$\rangle$}} & \multicolumn{1}{c}{\multirow{2}{*}{QMAP}} & \multicolumn{2}{c}{ML-QLS} \\
 & \multicolumn{1}{c}{} & \multicolumn{1}{c}{} & \multicolumn{1}{c}{} & \multicolumn{1}{c}{} & sRefine & V cycle \\ \hline
\multirow{11}{*}{Eagle} & adder\_28 & 87 & 123 & 127 & \textbf{54} & \textbf{45} \\
 & adder\_64 & 261 & 309 & 351 & \textbf{156} & \textbf{156} \\
 & adder\_118 & 609 & 690 & 685 & 735 & \textbf{510} \\
 & cat\_65 & 121 & 12 & 5 & \textbf{0} & \textbf{0} \\
 & dnn\_33 & 95 & \textbf{69} & 83 & 75 & 71 \\
 & ghz\_40 & 62 & 5 & \textbf{0} & \textbf{0} & \textbf{0} \\
 & isling\_34 & 22 & 38 & 58 & \textbf{0} & \textbf{0} \\
 & isling\_98 & 133 & 122 & 62 & \textbf{0} & \textbf{0} \\
 & knn\_31 & 59 & 82 & 88 & 64 & \textbf{54} \\
 & knn\_67 & \textbf{133} & 177 & 211 & 167 & 143 \\
 & wstate\_33 & 85 & 6 & \textbf{0} & \textbf{0} & \textbf{0} \\ \hline
\multicolumn{2}{c}{Geo. Ratio} & 8.10 & 4.66 & 3.25 & 1.09 & \textbf{1.00} \\ 
\hline \hline
\end{tabular}
\vspace{2pt}
\caption{SWAP count comparison for circuits from QASMBench benchmark suites on an IBM Eagle coupling graph.}
    \label{tab:qasm-eagle}
\vspace{-18pt}
\end{table}

\begin{figure}[t]
    \centering
    \includegraphics[width=0.95\linewidth]{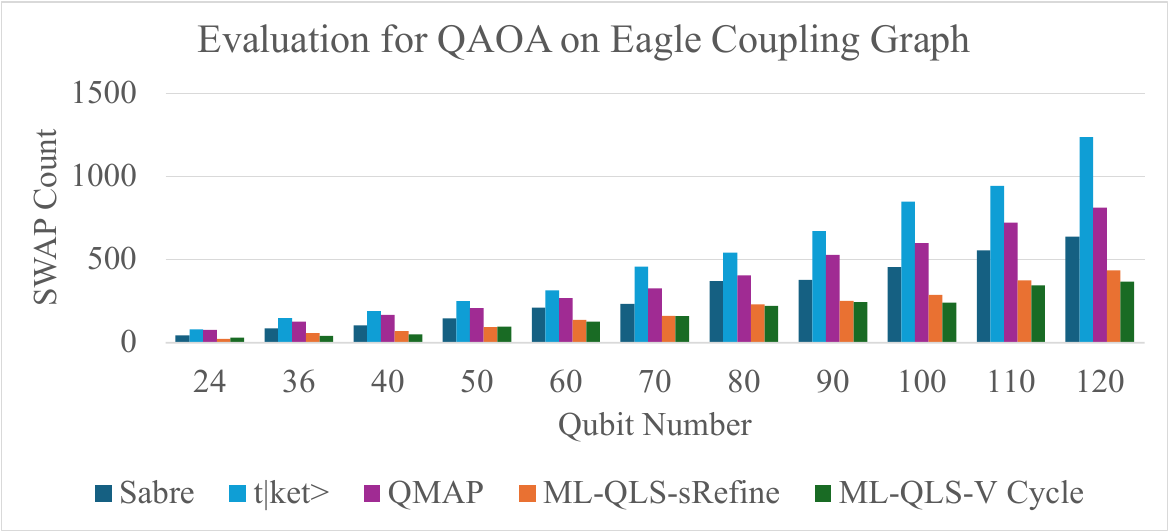}
    \caption{SWAP count comparison for QAOA circuits of size ranging from 24 to 120 qubits on an IBM Eagle coupling graph.}
    \label{fig:qaoa-eagle}
    \vspace{-12pt}
\end{figure}

\begin{figure}[htbp]
    \centering    \includegraphics[width=0.95\linewidth]{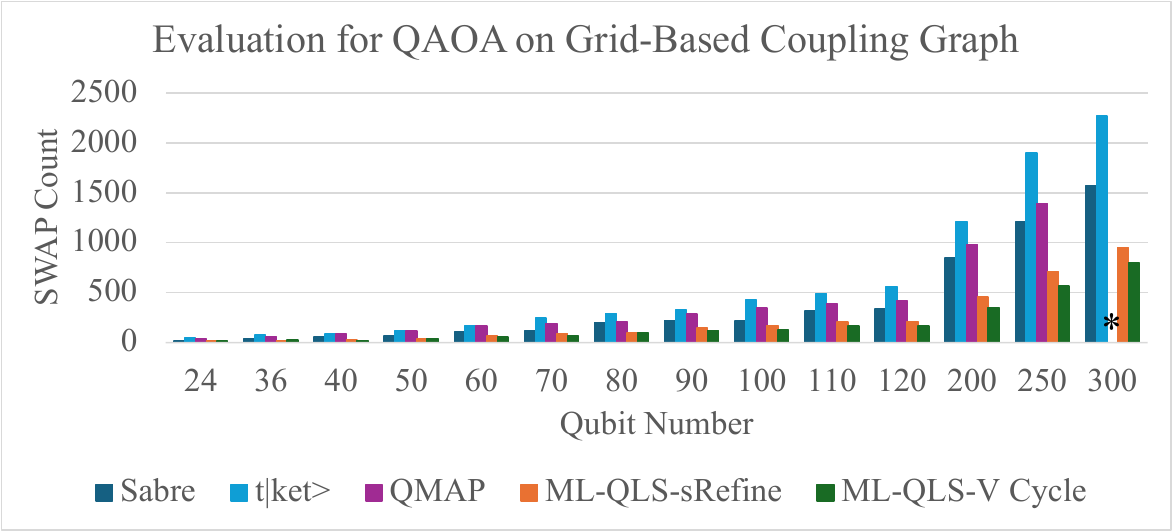}
    \caption{SWAP count comparison for QAOA circuits of size ranging from 24 to 300 qubits on grid-based coupling graphs.
    $*$ signifies that the bar of QMAP for QAOA-300 is missing because QMAP fails to return a solution within a timeout limit of 8 hours.}
    \label{fig:qaoa-grid}
    \vspace{-8pt}
\end{figure}


\begin{table}[]
\begin{tabular}{cllllll}
\hline \hline
\multirow{2}{*}{Device} & \multicolumn{1}{c}{\multirow{2}{*}{$D$}} & \multicolumn{1}{c}{\multirow{2}{*}{Sabre}} & \multicolumn{1}{c}{\multirow{2}{*}{t$|$ket$\rangle$}} & \multicolumn{1}{c}{\multirow{2}{*}{QMAP}} & \multicolumn{2}{c}{ML-QLS} \\
 & \multicolumn{1}{c}{} & \multicolumn{1}{c}{} & \multicolumn{1}{c}{} & \multicolumn{1}{c}{} & sRefine & V-Cycle \\
 \hline
\multirow{2}{*}{Sycamore} & 20 & 135 & 336 & 398 &\textbf{0} & \textbf{0} \\
 & 30 & 132 & 375 & 590 & \textbf{0} & \textbf{0} \\ \hline
\multirow{2}{*}{Grid 15$\times$15} & 20 & 1374 & 3636 & 5350 & 1396 & \textbf{1317} \\
 & 30 & 2073 & 5100 & 8145 & \textbf{1619} & \textbf{1619} \\ \hline
\multirow{2}{*}{Grid 20$\times$20} & 20 & 3426 & 9470 & 14286 & 4218 & \textbf{2972} \\
 & 30 & 4812 & 13194 & 20801 & 5375 & \textbf{3670} \\ \hline
\multirow{2}{*}{Grid 25$\times$25} & 20 & 8159 & 19320 & 29336 & 6418 & \textbf{4525} \\
 & 30 & 11131 & 26319 & 41942 & 9320 & \textbf{7483} \\
 \hline
\multicolumn{2}{c}{Geo. Ratio} & 4.20 & 10.82 & 16.20 & 1.19 & \textbf{1.00} \\
\hline \hline
\end{tabular}
\vspace{4pt}
\caption{The SWAP count measured by QUEKO circuits with an optimal depth $D$=20, 30 for Google Sycamore architecture and grid architectures with grid length 15, 20, and 25.}
    \label{tab:queko}
    \vspace{-18pt}
\end{table}

\vspace{8pt}
\subsection{Evaluation on QASMBench Circuits}
\label{subsec:eval_qasm}
The results for circuits from QASMBench benchmark suite on the grid architectures are reported in Table~\ref{tab:qasm-grid}, and the results on the Eagle architecture are shown in Table~\ref{tab:qasm-eagle}.
For ML-QLS, initial and final QLS results from each stage are provided. 
Note that the initial result is obtained from the standalone version of sRefine as described in Section~\ref{subsec:standalone_tool}.
V cycle can reduce SWAP gates by 9\% compared to sRefine, and demonstrates more significant improvement for the larger circuits.
This trend underscores the effectiveness of the multilevel framework, showing that the coarsest-level solution provides good guidance for subsequent refinement. 
In addition, V cycle achieves larger improvement on grid architectures than the Eagle architecture does (up to 46\%), 
and we posit the reason is that the dense connectivity in coupling graphs alleviates the inaccuracy in cost estimation in the multilevel framework.
As discussed in Section~\ref{subsubsec:challenge3}, the estimation of the cost increase for coupling graphs with sparse connectivity is difficult since we may not have all-to-all connectivity between coarser qubits.
In summary, ML-QLS achieves an average of 69\% SWAP reduction for grid architectures compared to QMAP, and 39\% reduction for Eagle architectures compared to QMAP.
In terms of scalability, ML-QLS takes less then 5 minutes to solve each problem while other tools takes less than 1 minute. 
The runtime overhead comes mainly from coarsest-level solving and A*-based SWAP insertion, which may be reduced by trading the solution quality via tuning the hyper parameters.

\subsection{Evaluation on QAOA Circuits}
\label{subsec:eval_qaoa}
Figure~\ref{fig:qaoa-eagle} and~\ref{fig:qaoa-grid} demonstrate SWAP count comparison for QAOA circuits across the Eagle architecture and the grid architectures with grid length set to be $\lceil\sqrt{|Q|}\rceil$.
The performance of V cycle is 9\% and 18\% better than sRefine for the Eagle architecture and grid architectures, respectively.
We observe that V cycle can perform better on the grid architectures,
which is in accordance with our observation in Section~\ref{subsec:eval_qasm}.
Moreover, comparing the performance gap for sRefine and V cycle on the same architecture, we observe that the gap is enlarged when the problem sizes increase,
showing that our multilevel framework is more powerful when dealing with large instances.
In summary, ML-QLS achieves an average of 42\% improvement with a maximal 53\% improvement and a minimal improvement of 36\% against Sabre for the Eagle architecture.
For grid architecture, the average improvement is 50\% with a maximal 59\% improvement and a minimal improvement of 39\%.
For the solving time, compiling each circuit takes less than 1 hour.

\subsection{Evaluation on QUEKO Circuits}
The evaluation results for QUEKO circuits are demonstrated in Table~\ref{tab:queko}.
In this experiment, we evaluate both methods on QUEKO circuits with an optimal depth 20 and 30 for Google Sycamore architecture and grid architectures with grid length 15, 20, and 25.
With InitialMapper, our tool can effectively find a SWAP-free mapping and achieve optimal circuit depth for circuits with less than 100 qubits, e.g., ising circuits.
Thus, we can always obtain the optimal results for Google Sycamore architecture.
On the other hand, due to the randomness and heuristics characteristics in Sabre, Sabre demonstrate large optimality gap.
In terms of the solving time, ML-QLS can return a solution within two minutes for all 54-qubit circuits.
For large instances, we do not invoke InitialMapper due to scalability concern for an SMT solver.
According to the table, V cycle produces higher quality results than sRefine,
demonstrating the effectiveness of the multilevel framework.
Compared to Sabre, ML-QLS can achieve an average of 31\% SWAP reduction.
Although ML-QLS achieves smaller SWAP count compared to others, it did not reach zero-SWAP solutions for large circuits.
As finding a SWAP-free solution is equivalent to solving the subgraph isomorphism problem, which is NP-complete, we will not have an efficient algorithm for this task,
indicating that there are further research to be done to search for more scalable and optimal QLS methods.
\color{black}

%% file: section/6_conclusion.tex
\section{Conclusion}
\label{sec:conclusion}
In this paper, we propose the first multilevel quantum layout tool, ML-QLS. 
This tool includes sRefine, which is an effective heuristic QLS tool that incorporates a novel cost function to capture the SWAP cost, clustering strategy, and efficient refinement.
Our experimental results demonstrate that ML-QLS has a 69\% performance improvement over the leading heuristic QLS tool for large circuits.
ML-QLS showcases the efficacy of multilevel frameworks in quantum applications, offering valuable insights for researchers exploring hierarchical approaches.
Its accomplishments highlight the significance of iterative, multilevel strategies in optimizing quantum compilation processes. ML-QLS sets a precedent for advancing quantum layout tools and strategies, paving the way for future developments in quantum computing optimization.

\section{Acknowledgements}
The authors would like to thank Bochen Tan, Hanyu Wang, Jason Kimko, Yunong Shi, Eric Kessler, and Yaroslav Kharkov for useful discussion on QLS and valuable comments on the manuscript.
This research is supported in part by the National Science Foundation Award 2313083 and Amazon under the Science Hub program.